\documentclass{aa}

\usepackage{astron}
\usepackage{graphics}
\usepackage{epsfig}

\begin {document}

   \thesaurus{ 12.04.3}

\titlerunning{Surface Brightness Fluctuation}

\title{K--band versus I--band Surface Brightness Fluctuations as distance indicators}

\author{S.Mei \inst{12}, P.J. Quinn \inst{2}, D.R. Silva \inst{2}}

\institute{ 
 Observatoire Midi--Pyr\'en\'ees, 14 ave. E. Belin, 31400 Toulouse, France; mei@ast.obs-mip.fr
\and
European Southern Observatory, Karl-Schwarzschild-Strasse 2, 85748 Garching, Germany 
}

%\markboth

\date{Received February 10, 2000 \/ Accepted 14 March 2001 }

\maketitle

\begin{abstract}

We evaluate the method of optical and infrared Surface 
Brightness Fluctuations (SBF) as a distance indicator and its 
application on 8--m class telescopes,
such as the Very Large Telescope (VLT).
The novelty of our approach resides in the development of  
Monte Carlo simulations
of SBF observations incorporating realistic elliptical
galaxy stellar population models, the effects induced 
by globular clusters and background galaxies, instrumental
noise, sky background and PSF blurring. 
We discuss, for each band and
in different observational conditions, the errors on distance 
measurements arising from stellar population effects, data treatment 
and observational constraints.
With 8--m class telescopes, one can extend I--band SBF 
measurements out to 6000--10000 km/s. 
Integration times in the K--band are too expensive from the ground, 
due to the high infrared background for large-scale distance determination projects. Nevertheless ground--based K--band measurements  
are necessary to understand stellar 
population effects on the SBF calibration, and to prepare 
future space--based observations, where this band is
 more efficient.


\end{abstract}

\section{Introduction}

Distance measurements are the key to understanding many central 
questions in astronomy.  Correct distance measurements permit
one to map the matter distribution in the universe, 
study its dynamics,  and to constrain certain cosmological 
parameters, such as $H_o$ and $\Omega_o$.
Beyond the limits reached by the Cepheid distance scale 
($\sim$~2000 km/s), the method of Surface Brightness Fluctuations (SBF) 
is one of the most accurate currently known, with a claimed error 
on single galaxy distances of $8-10\%$.  The approach is  
based on simple concepts.  It can in principle be used to 
calibrate other methods, such as Tully-Fisher, $D_n-~\sigma$ and 
Supernovae I \cite{ton97}, that extend beyond the limit
of SBF detection.  It is therefore important to understand the method's 
limitations in terms of distance and to identify the various 
sources of error affecting SBF measurements. 

\noindent Recently, SBF have been successfully used to measure 
distances to galaxies with radial velocities up to 4000 km/s with
ground--based telescopes, and up to 7000 km/s in optical bands using the 
Hubble Space Telescope \cite{sod95,sod96,aj97,ton97,tho97,lau98,pah99,ton00,bla01,mei00,ton01}.
A recent review of the method can be found in Blakeslee et al. (1999).
The technique was introduced by Tonry and Scheineder (1988) and 
is based on a simple concept: the Poisson distribution of 
unresolved stars in a galaxy produces fluctuations in each pixel 
of the galaxy image. The variance of these fluctuations is proportional 
to the square of the flux of each star (f$\propto 1/d^2$) times the 
number of stars per pixel(n~$\propto~d^2$). 
While the mean flux per pixel does not depend on distance, the variance 
is inversely proportional to the square of the galaxy distance.
The SBF amplitude is defined as the variance normalized to the mean 
flux of the galaxy, which then indicates the flux-weighted average stellar 
flux.  This implies that the brightest stars (in evolved populations, 
the red giant branch) contribute most to the signal. 

\noindent The SBF signal is convolved with the point spread function 
(PSF) created by the telescope optics and atmospheric seeing. This 
permits one to distinguish the galaxy fluctuations from white noise
fluctuations in the image (e.g., photon shot noise, read--out noise 
and dark current), because in the Fourier space white noise fluctuations have
a flat spectrum.
At the same time, the fluctuations are contaminated by the signal 
from external sources, such as background galaxies, globular clusters 
and foreground stars, that are also convolved with the PSF.
For these reasons, obtaining good SBF measurements requires good seeing 
and sufficiently long integration times to both extract 
the signal from photon shot noise, and to detect external 
sources and correctly estimate their luminosity function;
the latter then serves to estimate and remove the
contribution of faint, undetected sources from the SBF amplitude.

\noindent In addition, stellar populations vary from elliptical 
to elliptical, and the absolute amplitude of the fluctuations 
depends on these variations. This implies that the SBF amplitude 
is not {\em per se} a standard candle at each wavelength, but rather 
that it must be calibrated as a function of stellar population.
Tonry and his collaborators have found an empirical calibration for 
the absolute magnitude of I--band SBF, via a linear relation with 
(V--I) color that demonstrates a universal zero point and slope \cite{ton00,fre00}.
This calibration is supported by theoretical predictions 
from Worthey's stellar population models \cite{wor93,wor94,jen98}. 
When making optical SBF measurements, it is thus necessary to 
have both V and I-band images of the observed galaxies, to 
obtain the precise (V~--~I) photometry necessary for the calibration.

\noindent In recent years the method has been extended to infrared 
bands for several reasons \cite{lup93,pah94,jen96,jen98,jen99,mei01a,mei01b,jen00}. 
Firstly, in these bands the contrast between the brightest red 
giants and the underlying stellar population is more extreme than 
in the optical, producing larger SBF at any given distance.
Secondly, the color contrast between the SBF and point sources 
(globular clusters and background galaxies) is larger than in 
the I--band.  Moreover, at ground facilities, the K--band seeing 
is intrinsically better than I--band seeing, which further enhances 
the SBF contrast.

\noindent Relative to the theoretical predictions 
\cite{wor93,buz93,wor94,jen98,mei99,liu99,liu00,bla01}, 
K-band SBF absolute amplitudes are predicted to have only a
weak dependence on (V--I) and a potentially larger scatter, 
because the effects of age and metallicity are less degenerate 
than in the I-band.  Ellipticals measured to date 
\cite{lup93,pah94,jen96,jen98,jen99,jen00} show an almost constant 
K-band SBF absolute amplitude, suggesting that these fluctuations 
may provide a good standard candle.
These arguments lead one to believe that, in the infrared, it is possible 
to make SBF measurements to a greater distance than in the optical, 
and that in the future, most SBF measurements will be taken at this 
wavelength.  On the other hand, the ground--based  background level in infrared bands 
is higher than in the optical, making detection of extended 
sources more difficult.  Jensen and his collaborators have 
measured K-band SBF out to 7000~km/s from 
ground--based telescopes \cite{jen99,jen00}.
Their integration times did not permit them to obtain a high 
signal--to--noise ratio, and they also found residual spatial 
variations that contaminated their SBF measurements.
The low signal--to--noise ratio of most current observations, 
the limited size of the sample and the low (V--I) coverage of 
the Jensen et al. (1998) sample ((V~--~I) $>$ 1.15)
prevent us from being entirely confident in  
the precision of K-band SBF measurements as a standard candle.
Moreover, some of the ellipticals with SBF distance measurements in 
both I and K-bands show a discrepancy between the two 
\cite{pah94,jen96,jen98,mei01a,mei01b}, suggesting that either the 
measurements are affected by systematic errors (i.e., insufficient 
signal--to--noise), or by the presence of stellar populations 
that can alter the measurement and that are not taken properly 
into account in the present calibration.
Anomalously bright cool stellar populations may already have been detected  in some ellipticals
\cite{els92,fre92,pah94,sil98,lup93,jen96,jen98,mei01a,mei01b}. 

\noindent In this paper we examine the accuracy and usefulness of the SBF 
method by studying the different sources of error inherent in the method. 
We wish to quantify the error budget for SBF measurement due to: a) stellar 
population effects; b) observational constraints; and c) data processing.
This study permits us to better understand the potential and the limitations 
of the method when applied in different wave bands and at different distances.
We have simulated galaxies in the I and in the K~-~band using theoretical models by Bruzual 
\& Charlot \cite{bru93,cha96,bru00,liu00} to study the effects of population 
variations in the infrared. To I-band and K-band models we have added 
external sources and then extracted distance measurements after varying 
the SBF signal--to--noise ratio, the seeing, and completeness magnitude 
for point sources detection. 
In \S 2 we discuss the role of different stellar populations.
We present our analysis of the observational constraints and data 
treatment in \S 3, and discuss the results in \S 4.

\section{\bf SBF and stellar population effects in the K-band}

\subsection{Overview}

The attraction of a high intrinsic SBF amplitude in the infrared may be 
negated by the effect of stellar population variations. By studying 
stellar population effects, our aim is to understand the error on the 
empirical calibration in the K-band from a theoretical point--of--view, and 
the error that arises from the presence of different stellar populations.
This can be seen as a two--fold problem. On one hand, if the dispersion in
K--band absolute magnitudes, or in its dependence on galaxy colors, is small 
enough, we will be able to make distance measurements with reasonable 
accuracy. On the other hand, if different stellar populations have 
significantly different absolute magnitudes, we might be able to distinguish 
populations of different age or metallicity.
%Table -------------------
\begin{table*}

\begin{flushleft}
\begin{tabular} {ccccccccc} \hline \hline
Age&z&$\overline{M}_K$&V-I&V-K&$\overline{M}_K$&V-I&V-K\\ 
&&(SP)&(SP)&(SP)&(EMP)&(EMP)&(EMP)\\
(Gyr)&&(mag)&(mag)&(mag)&(mag)&(mag)&(mag)\\ \hline
3&0.05&-6.33&1.19&3.35&--&--&--\\ 
&0.02&-5.83&1.09&2.94&-5.85&1.11&2.97\\
&0.008&-5.39&1.03&2.59&--&--&--\\ 
&0.004&-5.14&0.96&2.34&--&--&--\\ \hline

5&0.05&-6.18&1.27&3.51&--&--&--\\ 
&0.02&-5.62&1.14&3.00&-5.63&1.15&3.04\\
&0.008&-5.25&1.04&2.62&--&--&--\\ 
&0.004&-5.07&1.00&2.41&--&--&--\\ \hline

7&0.05&-6.17&1.32&3.61&--&--&--\\ 
&0.02&-5.48&1.18&3.04&-5.49&1.19&3.12\\ 
&0.008&-5.50&1.11&2.78&--&--&--\\ 
&0.004&-5.02&1.05&2.51&--&--&--\\ \hline

12&0.05&-5.90&1.37&3.67&--&--&--\\ 
&0.02&-5.50&1.25&3.25&-5.50&1.27&3.30\\ 
&0.008&-5.34&1.17&2.91&--&--&--\\ 
&0.004&-4.83&1.10&2.60&--&--&--\\ \hline

15&0.05&-5.77&1.39&3.71&--&--&--\\ 
&0.02&-5.48&1.28&3.30&-5.48&1.30&3.34\\
&0.008&-5.19&1.20&2.94&--&--&--\\ 
&0.004&-4.73&1.12&2.63&--&--&--\\ \hline \hline

\end{tabular}

\end{flushleft}

\caption{K-band absolute magnitudes from Salpeter (SP) and empirical (EMP) IMF Bruzual and Charlot models} \label{tab-cha}
\end{table*}
%--------------------------

\noindent Two recent papers have explored the effects of changes in 
stellar populations on the SBF amplitudes, Liu et al. (2000) and Blakeslee et al. (2001).  While models 
based on Bruzual \& Charlot (2000) \cite{liu00} predict a brightening 
of the amplitudes of K--band SBF absolute fluctuations for 
redder populations, independent models from Blakeslee et al. (2001) 
predict a dimming of the K--band absolute fluctuations for redder 
populations. At present, observations are not extensive enough in  
luminosity and color range to discriminate between the two predictions. 
We focus in this paper on Bruzual \& Charlot (2000) models \cite{liu00} that we have used in our Monte Carlo simulations.

\noindent Our simulations use single burst Bruzual and Charlot 
\cite{bru93,cha96,bru00,liu00} stellar population models, based on Padova tracks and semi-empirical SEDs \cite{liu00}.
Further details on these models can be found in Liu et al. 
(2000), with a wide range of predictions for SBF in different observing 
bands and for different ages and metallicities. 
Liu et al. (2000) also used single--burst population models. We add in this work a section on composite models. 
The I and K filter used in our work correspond to the filters of the instruments FORS1 and ISAAC at the ESO 8.2m VLT/UT1 telescope (http://www.eso.org/instruments/fors1/, http://www.eso.org/instruments/isaac/). In the I--band we used a Bessel filter.
 In the K--band, we used a $K_{short}$ filter, according to the VLT ISAAC 
instrument.
The difference between the standard K filter and $K^{'}$,  which 
is the filter adopted by Worthey in his SBF predictions, is $ \leq 0.03$ 
mag \cite{jen98}.  Since the $K_{short}$ filter is between these two filters, our 
results can be compared with Worthey's without making magnitude corrections.
Bruzual and Charlot models differ from Worthey models in their stellar 
evolution prescriptions \cite{cha96}. In particular, they differ in how 
the stars evolve once they leave the Main Sequence and in the way 
colors are assigned to stars in specific positions in the theoretical 
color--magnitude diagram.  Stellar population models have well--known 
uncertainties, typically of $\approx$ 0.03 mag in B--V colors, 
amounting to 0.1 - 0.14 mag in V--K colors.

\noindent In their latest models, Bruzual and Charlot have refined the 
prescription for cool RGB (Red Giant Branch) and AGB (Asymptotic Giant 
Branch) stars in their models. In particular, they use an improved 
color--temperature calibration for cool stars \cite{fea96}. Moreover 
they account for the evolution of M stars into C stars near the tip of 
the AGB, and its dependence on stellar mass and metallicity. The 
prescription is semi--empirical and based on models and observations 
of evolved stars in the Galaxy, the LMC and the SMC \cite{liu00}.

%Figure--------------------
\begin{figure*}[!htf]
\centerline{{\psfig{file=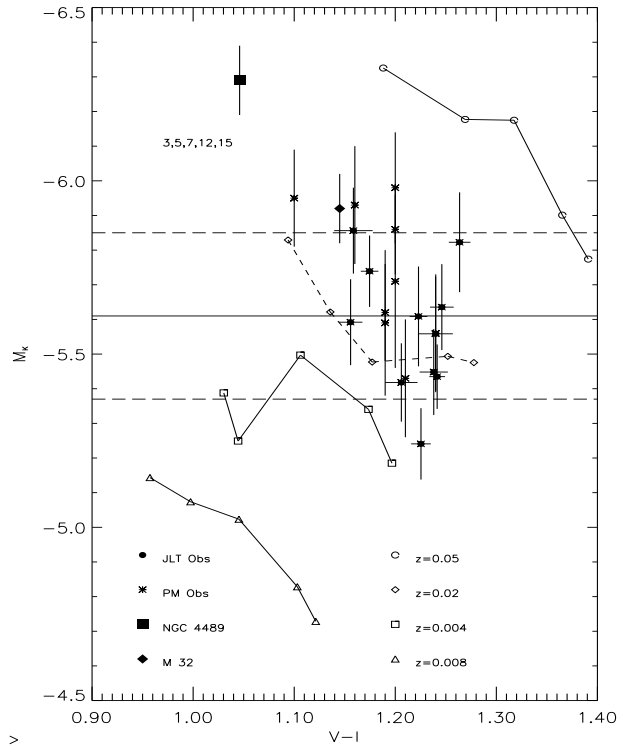,width=8cm,height=10cm}}}

\caption {We plot K-band SBF from recent observations versus Bruzual 
and Charlot(2000) single burst stellar population models predictions. 
The filled circles are data from Jensen et al. (1998) (JLT), while
the asterisks are the data from Pahre \& Mould (1994) 
(PM). The filled square is NGC 4489 from Mei et al. (2001a) and the filled diamond is M~32 from Luppino \& Tonry (1993).  Ages increase from left to right, taking on values of 3, 5, 7, 12 and 15 Gyrs. Solar metallicity models are plotted as triangles, 
twice solar metallicity as empty circles, 40\% solar metallicity as 
diamonds and 20\% solar metallicity as squares.}
\label{fig-cha1}
\end{figure*}
%--------------------------

\subsection{Single--age, single--metallicity stellar populations}

\noindent We considered ages equal to 3, 5, 7, 12 and 15 Gyr, 
and metallicities Z= 0.004 (20\% solar),  0.008 (40\% solar), 0.02 
(solar) and 0.05 (twice solar).
In the case of solar metallicity, we had a choice between a 
library of empirical stellar spectra \cite{lej97} and one of 
theoretical model atmospheres \cite{gun83,per83,flu94}. 
For the adopted  initial mass function, solar metallicity and 
formation redshift $z_f=5$, the models provide good fits to 
the spectral features and colors of nearby elliptical and S0 galaxies, 
from the ultraviolet to the infrared  \cite{kau98}.  

\noindent We summarize the predictions of these models that are 
important in the context of our simulations in  Figure~\ref{fig-cha1}  
and in Table~\ref{tab-cha}. 
Additional model predictions and an interesting discussion of the 
models can be found in Liu et al. (2000).  The comparison with Blakeslee et al. (2001) can be found in that study.

\noindent  The SBF amplitude was calculated as: 
\begin{equation}
\sigma^2_{SBF}=\frac{ \sum_{i=1}^{N_{pop}} n_{i}f^2_i } {\sum_{i=1}^{N_{pop}} n_i f_i }, \label{eq:pop1}
\end{equation}
\noindent where the sum is taken over the different species of 
the underlying stellar population, with the assumption that the 
same stellar population is present in each pixel of the galaxy. 
We first calculated the amplitude of K-band fluctuations in single 
population models. 

\noindent In Figure~\ref{fig-cha1} we plot Bruzual and Charlot models with 
a Salpeter IMF in comparison to observed SBF K-band absolute magnitudes. 
The asterisks are the  data with errors from Pahre \& Mould (1994) (PM) 
and the filled circles are the high signal--to--noise data (as defined 
by Jensen et al. (1998)) with errors from Jensen et al. (1998) (JLT).

\subsection{Composite Stellar Populations}

\noindent We have also explored the case of composite populations by mixing two different 
stellar populations to study simple combinations of age and 
metallicity, for Bruzual \& Charlot (2000) models.  

\begin{figure}

\centerline{\psfig{file=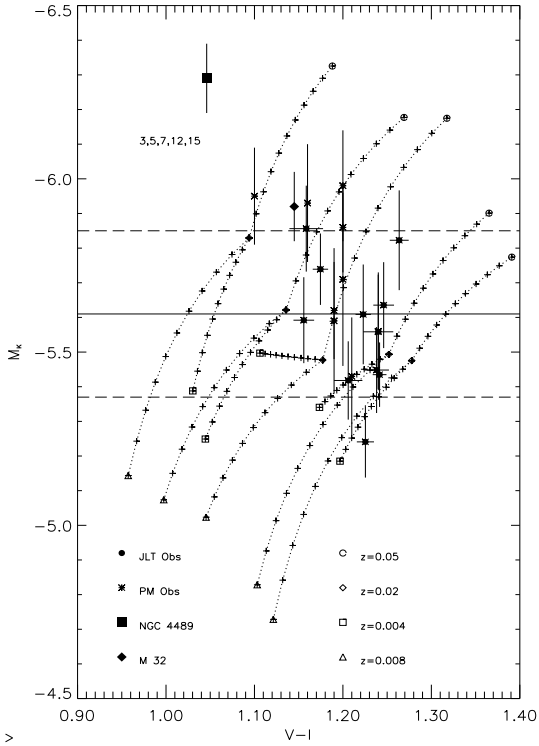,width=7cm,height=10cm}}

\caption{Composite stellar population effects. 
We plot K-band SBF from recent observations versus Bruzual and 
Charlot(2000) single burst  model predictions, shown as the open symbols.  We consider
here the case of a composite population consisting of two
identical age sub--populations with different metallicities. 
The filled circles are 
the data from Jensen et al. (1998) (JLT), while the asterisks are the 
data from Pahre \& Mould (1994) (PM). The filled square is NGC 4489 from Mei et al. (2001a) and the filled diamond is M~32 from Luppino \& Tonry (1993).  
For the single--burst models, ages increase from left to right and take on values of 3,5, 7,12 and 15 Gyrs.  
For the composite populations, the main sub--population is fixed at solar metallicity, and it's
position on the plot is indicated by the five diamonds (one for each
age).  From each of these points emanates three dotted curves,
one for each metallicity of the second sub--population; each of
these dotted lines terminates in a symbol indicating the
metallicity of the second population.  Variations along
the dotted lines show the effect of changing the abundance
of the second population.  We measure this abundance
by the fractional contribution of the second population to
the total luminosity; the crosses indicate 10\% steps in
luminosity contribution.  The horizontal lines show the mean 
(solid line) and the two standard deviation variation 
(long dashed lines) of the observed $\overline{M}_K$.} \label{fig-cha2}
\end{figure}

\noindent In Figure~\ref{fig-cha2} we compare the
observations to Bruzual and Charlot models with
composite stellar populations, where each composite 
stellar population is a mixture of two identical--age 
populations of different metallicities.
To a main population of fixed age (3, 5, 7, 12, and 15 Gyrs) 
and solar metallicity ($Z=0.02$), we have added a second 
population with, respectively,
200\% (open circle), 40\% (open triangles) and 20\% 
(open squares) solar metallicity. We quantify the 
presence of this second population in terms of its fractional 
contribution to the total light.  Thus, we see a trio
of dotted lines (one for each metallicity of the second
population) emanating from each of the five diamonds 
(one for each age of the main population).  The abundance 
of the second population varies along each dotted line, and the
crosses indicate steps of 10\% variation in the luminosity
contribution of this second population (0\% at the
position of the diamond, and 100\% at the position of
the symbol corresponding to the metallicity of the
second population).      
The fluctuation colors of these simple 
composite models are all compatible with the current observations 
($\overline V - \overline I > 1.95$ and $\overline I - \overline K > 3.65$ 
\cite{bla01}). 

\noindent We conclude from these models  that
while young populations with high metallicities tend to increase the 
amplitude of the fluctuations, low metallicities lower their value. 
The K--band SBF amplitude becomes higher for redder populations.  

\noindent At present, some galaxies with fluctuations higher than 
the Jensen et al. (1998) mean have been observed 
\cite{lup93,pah94,jen96,jen98,mei01a,mei01b}, while there are no cases of 
galaxies observed with anomalously low fluctuations.

\begin{figure}

\centerline{\psfig{file=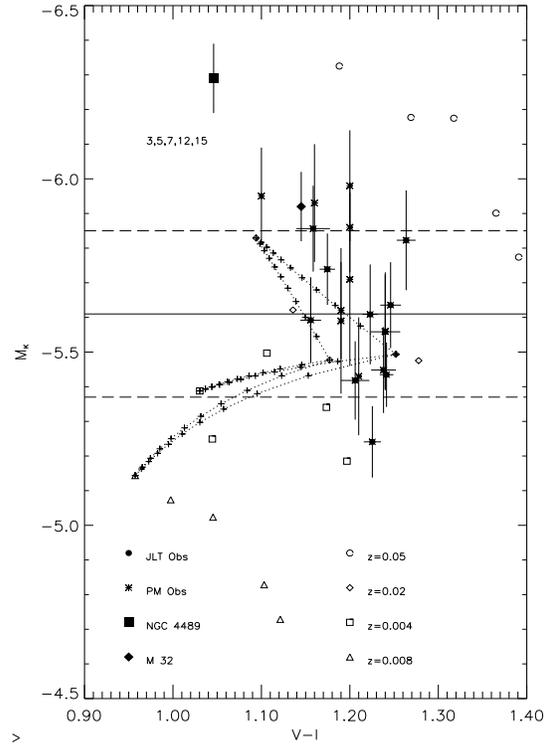,width=7cm,height=10cm}}

\caption{Composite stellar population effects. 
We plot K-band SBF from recent observations versus Bruzual and 
Charlot(2000) single burst  model predictions, shown as the open symbols.  We consider
here the case of a composite population consisting of one
old age (12~Gyr and 15~Gyr) and one 3~Gyr  sub--population with,
respectively, solar and solar and sub--solar metallicities. 
The filled circles are 
the data from Jensen et al. (1998) (JLT), while the asterisks are the 
data from Pahre \& Mould (1994) (PM). The filled square is NGC 4489 from Mei et al. (2001a) and the filled diamond is M~32 from Luppino \& Tonry (1993).  
For the single--burst models, ages increase from left to right and take on values of 3,7,12 and 15 Gyrs.  
For the composite populations, the main sub--population is fixed at solar metallicity and age 12~Gyr and
15~Gyr.  From each of these two points emanates three dotted curves,
one for each solar and subsolar metallicity of the second sub--population,
that is fixed at age 3~Gyr.  Variations along
the dotted lines show the effect of changing the abundance
of the second population.  We measure this abundance
by the fractional contribution of the second population to
the total luminosity; the crosses indicate 10\% steps in
luminosity contribution.  The horizontal lines show the mean 
(solid line) and the two standard deviation variation 
(long dashed lines) of the observed $\overline{M}_K$.} \label{fig-cha3}
\end{figure}

\noindent We have also studied the case of a composite model
consisting of an old (12 and 15 Gyrs) 
solar metallicity population mixed with a 3 Gyr population 
with metallicities lower than or equal to solar (Figure~\ref{fig-cha3}). 
The average K--band SBF amplitude for the old solar 
metallicity population is -5.49 mag. A one sigma detection 
of the 3 Gyr old population with respect to the Jensen et al. (1998) 
mean is a measurement of K--band SBF fluctuations brighter than 
-5.73; at two sigma, the K--band SBF fluctuations are brighter 
than -5.85. A composite 12/15 Gyr plus a 3 Gyr solar metallicity 
population will produce K--band SBF brighter than -5.73  
when 50\% of the total light is contributed by the younger
population; it is never brighter than -5.85.   A 
composite population of 12/15 Gyr solar metallicity plus 
a 3 Gyr at half solar metallicity population will never 
produce K--band SBF brighter than -5.73.
Now consider a composite consisting of 
a 15 Gyr population at twice solar metallicity mixed
with a 3 Gyr population at solar metallicity; this
will produce K--band SBF between -5.77 and -5.82, i.e.,
between 1 and 2 sigma off the mean.
Finally, single--burst populations with twice solar 
metallicity all have predicted K--band SBF brighter than -5.77. 
These predictions differ from Blakeslee et al. (2001) stellar 
population model predictions, where amplitudes brighter than 
-5.73 are predicted for composite populations with low (V--I) 
colors.  
 
\noindent In conclusion, SBF amplitudes higher than the Jensen et al. 
(1998) mean are consistent with both a high metallicity stellar 
population and the presence of an intermediate age population. 
However, as pointed out by Liu et al. (2000), if K--band SBF are
mainly sensitive to metallicity variations, they can be used in combination
with age sensitive observables to break the age--metallicity degeneracy in
elliptical galaxies. The predictions from Blakeslee et al. (2001) do
not show this possibility.
The NGC 4489 high fluctuations though are not consistent with any of these models (for discussion see Mei et al. 2000b). 

\noindent To discriminate between the currently available models, the sample of ellipticals with accurate K--band SBF measurements needs to be extended to (V--I) $<$ 1.15.  In general accurate K--band SBF are only available for galaxies with (V--I) $>$ 1.15. In that color range, our predicted SBF magnitudes and Liu et al. (2000) predicted magnitudes (both based on Bruzual \& Charlot (2000) models) as well as  Blakeslee et al (2000)  predicted magnitudes are consistent with each other and the available data. 
In our Monte Carlo simulations, we choose  single-burst 
Bruzual \& Charlot (2000) models closer to this mean 
and a bright magnitude. In fact, distance measurements at large distance in
the K--band will be made on luminous and red galaxies.
%Figure--------------------

%Figure--------------------

\section{Monte Carlo simulations of SBF measurements}

In this section, we study errors introduced into SBF measurements 
by non--ideal observing conditions and data treatment.
Our aim is to quantify the limits of the SBF method when measuring 
galaxies at large distance using large ground--based telescopes, 
such as the VLT, in both the I and the K--bands. 

\noindent To achieve this, we have isolated the different causes of 
error on the measurements and have studied their contribution 
as a function of distance and galaxy magnitude. 
As observational constraints, we have considered detector efficiency, 
detector scale and the seeing of the observations. 

\noindent Concerning data treatment, we have studied the error 
induced by the fitting of the galaxy power spectrum used to 
extract the SBF amplitude, both with and without contributions 
from external sources.

\noindent We then compare the results from the I and the K-bands, 
in order to understand the advantages and the limits of 
infrared and optical I-band SBF measurements. 

\subsection{Description of the simulations}
%FLOW CHART

We have simulated elliptical galaxies with a De Vaucouleurs 
surface brightness profile. 
The galaxies were constructed by randomly selecting stars from 
Bruzual \& Charlot(2000) theoretical luminosity functions.
Specifically, for these simulations we have chosen a model with a 
Salpeter IMF, solar metallicity (Z=0.02) and an age of 12 Gyr. 
This is the  model that best reproduces the observed average 
I and K--band SBF amplitudes. 
The absolute amplitude of the K-band SBF predicted from this 
model is $\overline{M}_{K} = -5.49$ mag, in agreement with the observed 
mean of $\overline{M}_{K} =  -5.61 \pm 0.12$ mag \cite{jen98}.
In fact, K--band observations at large distance will be mainly made on
bright and red elliptical galaxies, as for the Jensen et al. (1998) sample.
The absolute  amplitude of the I-band SBF predicted from this model 
is $\overline{M}_{I} = -1.03$ mag. The (V--I) color from the model is 
1.25, not too much different from the Jensen et al. (1998) sample, 
but the interesting point is that our simulations predict 
$\overline{M}_{I}$.  This value is lower than $\overline{M}_I$ from the Tonry 
et al. (2000) calibration for the same color.  All the other 
parameters being fixed, brighter I--band SBF will increase 
the signal--to--noise ratio ($P_0/P_1$, defined below), as well 
as the ratio of SBF to external 
source fluctuations. Thus, this lower value does not affect the results of this study.

\noindent After convolution with a Gaussian point spread 
function, we added a background of white noise with contributions
from the sky, galaxy shot noise, read--out noise and dark current. 
The sky magnitude was assumed to be 13 mag/arcsec$^2$ in K-band,
and 19.2 mag/arcsec$^2$ in I-band. These values are close to 
the sky brightnesses in grey time in Paranal.
For telescope parameters, we have considered zero magnitudes, 
read--out noise figures and dark currents, choosing values appropriate for  
the FORS1 and ISAAC instruments at the ESO 8.2m VLT/UT1 telescope (http://www.eso.org/instruments/fors1/, http://www.eso.org/instruments/isaac/). 
The read--out noise is thus set at 10 e$^-$ and the dark current 
$<$ 0.1 e$^-$/s for ISAAC, and respectively 6 e$^-$ and $<$ 0.001 e$^-$/s for FORS1. The pixel scale was assumed to be 0.15 $\arcsec$/pixel, 
which is the pixel scale of ISAAC (0.147 $\arcsec$/pixel), 
and a value between the two available choices for FORS1 
(0.1 $\arcsec$/pixel and 0.2 $\arcsec$/pixel). These parameters are 
summarized in Table~\ref{tab-4}.

%Figure--------------------

\begin{figure}[!htf]
\psfig{file=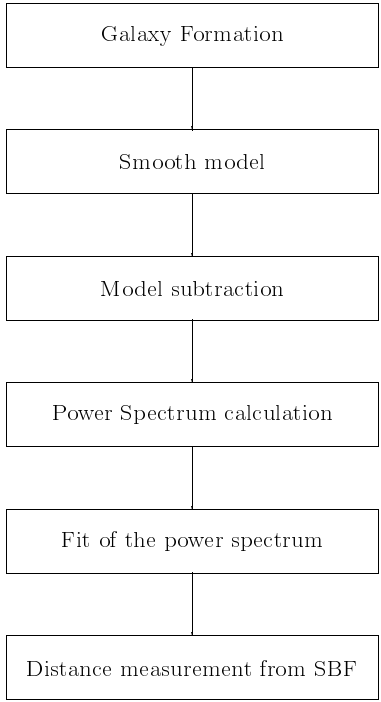,width=8cm,height=10cm}
\caption{Simulation procedure.} \label{fig-flow}
\end{figure}

%-------------------------

\noindent As a first step, the error on SBF fluctuations was 
calculated in the absence of external sources; this represents 
an ideal lower limit on the error in estimating the fluctuation power. 
The images were reduced using the standard SBF measurement procedure (see for example Tonry et al. 1997, 
Tonry et al.(2000), and Mei et al. 2000a for discussions).
The sky was first subtracted from the image, and then an average 
galaxy brightness profile was constructed by fitting isophotes 
with varying centers, ellipticity and orientation angle. 
We used the IRAF \footnote{The Image Reduction and Analysis Facility (IRAF) is distributed by the National Optical Astronomy Observatories} isophote package.
A galaxy model was thus obtained and then subtracted from the image.
Large--scale residuals were fitted and subtracted from the original image. 
The final residual image was then divided by the square root of 
the average galaxy profile, to normalize the fluctuations to 
the average galaxy flux and to make them constant over the image.
%Table -------------------

\begin{table*}

\begin{flushleft}
\begin{tabular} {|c|c|c|c|c|c|c|c|c|c|} \hline 

Seeing & Scale & $M_{V}$& $\overline{M}_{I}$& $\overline{M}_{K}$ & $m^I_{sky}$ &  $m^K_{sky}$ \\ \hline
($\arcsec$)& ($\arcsec$/pixel)&(mag)&(mag)&(mag)&(mag)&(mag) \\ \hline

0.6/1&0.15&-23&-1.03&-5.49&19.2&13\\ \hline

\end{tabular}

\end{flushleft}

\caption{Parameters used for the galaxy simulations.} \label{tab-4}
\end{table*}

%--------------------------
\noindent The resulting image was divided into concentric 
annuli, each with a radial extent of approximately 12 $\arcsec$. 
Over each annulus, we calculated the power spectrum as
the square of the modulus of the spatial Fourier transform.
The PSF power spectrum was found from the Fourier 
transform of the input Gaussian PSF.
This means that in our simulations we do not include errors that can come from an inaccurate estimation of the  PSF Fourier transform, because of lack of good high-S/N stars in both I and K--band images.

\noindent We then fit the image power spectrum with the following
functional form: a constant $P_{0}$ times the PSF power spectrum,
plus a constant term, $P_{1}$, representing the 
white noise power contribution:
\begin{equation}
E_{gal}= P_{0} \ E_{PSF} + P_{1}.
\end{equation}
\noindent From this we derive
\begin{equation}
P_{0} = \sum \sigma^2_{SBF} 
\end{equation}
\noindent and  

\begin{equation}
P_{1} = \sum \sigma^2_{ph} +\sum \sigma^2_{RON}.
\end{equation}

\noindent  where $\sigma^2_{SBF}$ represents the surface brightness 
fluctuations, $\sigma^2_{ph}$ the fluctuations due to photon 
shot noise, and $\sigma^2_{RON}$ the contribution from read out noise. 
The sum is taken over all non--zero pixels of the image.

\noindent The amplitude of the fluctuations is computed as:
\begin{equation}
\overline m= -2.5 log (P_{0}/t)+m_{0},
\end{equation}
\noindent  where $t$ is the assumed integration time and $m_{0}$ 
the magnitude for which one ADU/s is collected.

\noindent The steps followed in this procedure are summarized 
in Figure~\ref{fig-flow}.

\noindent The error on $P_0$ was estimated as follows:
The error calculated from a linear least squares fit, 
for chi--squared variations, assumes that the variables to be 
fitted are each extracted from a Gaussian distribution. 
This is not directly applicable to our situation, where
each power spectrum point has been calculated as
the square of the FFT of Gaussian variables (the pixels in the 
images in which we have on average more that 10 ADU). 
Our power spectrum values follow instead a 
chi--squared distribution \cite{sod95}.

\noindent In this case, the error calculated by a simple Gaussian least square fit does not represent the true fit errors. The order of magnitude of the actual error can be estimated by repeated Monte Carlo simulations. We simulate the same galaxy, with the same observational parameters, four hundred times and calculate the variance on the parameter $P_0$ obtained from the fits.

\noindent The galaxies were simulated with an absolute magnitude 
of $M_{V} = -23$ for distances from 10 to 130 Mpc.
The percentage error on the distance measurement was calculated as 
a function of the distance and of the radius of the considered 
annulus in the galaxy. 

\noindent The percentage error introduced by power spectrum 
fitting was calculated as a function of the signal--to-noise ratio, 
defined as: 
\begin{equation}
\frac{S}{N} \equiv  \frac{ \sum \tilde \sigma^2_{SBF}}{\sum \tilde
  \sigma^2_{ph} +\sum \tilde \sigma^2_{RON}}
\end{equation}

\noindent Where $\tilde \sigma^2_{SBF}$, $ \tilde \sigma^2_{ph}$ and $\tilde
  \sigma^2_{RON}$ are SBF, photon shot noise and read out noise fluctuations,
  normalised at the mean galaxy profile in each pixel, as defined in Tonry \&
  Schneider (1988).

\begin{figure*}

\centerline{\psfig{file=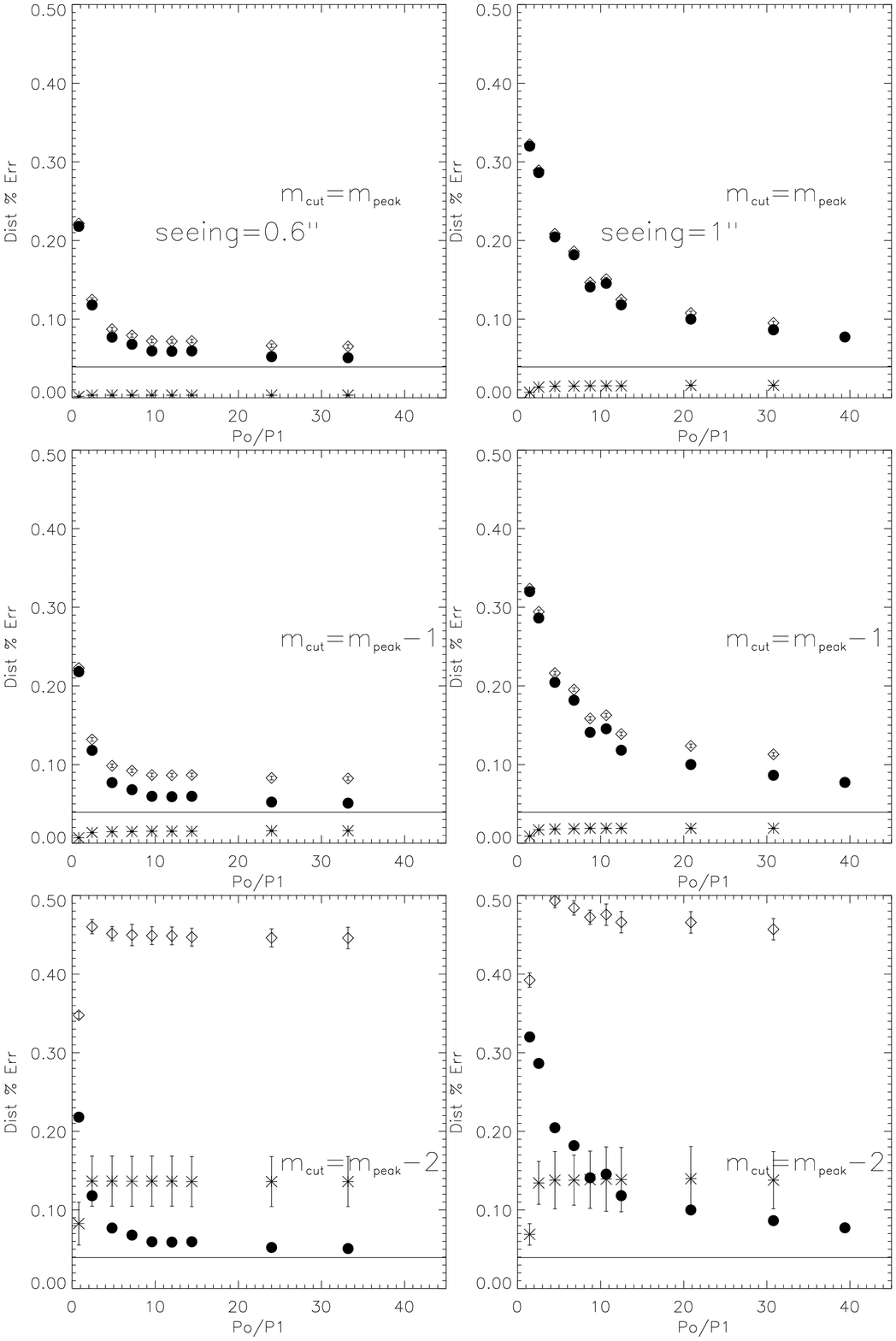,width=18cm,height=15cm}}

\caption{I--band error budget found from our simulations
for seeing = 0.6 \arcsec and 1 \arcsec.  The percentage error on distance 
is shown as a function of the signal--to--noise ratio. 
Diamonds represent the total error; filled circles, the error due only
to power spectrum fitting; and asterisks, the error due to external 
source detection and subtraction. The constant solid line 
is the percentage error induced by the zero point magnitude 
and SBF absolute magnitude.} \label{fig-perrI}

\end{figure*}

\begin{figure*}

\centerline{\psfig{file=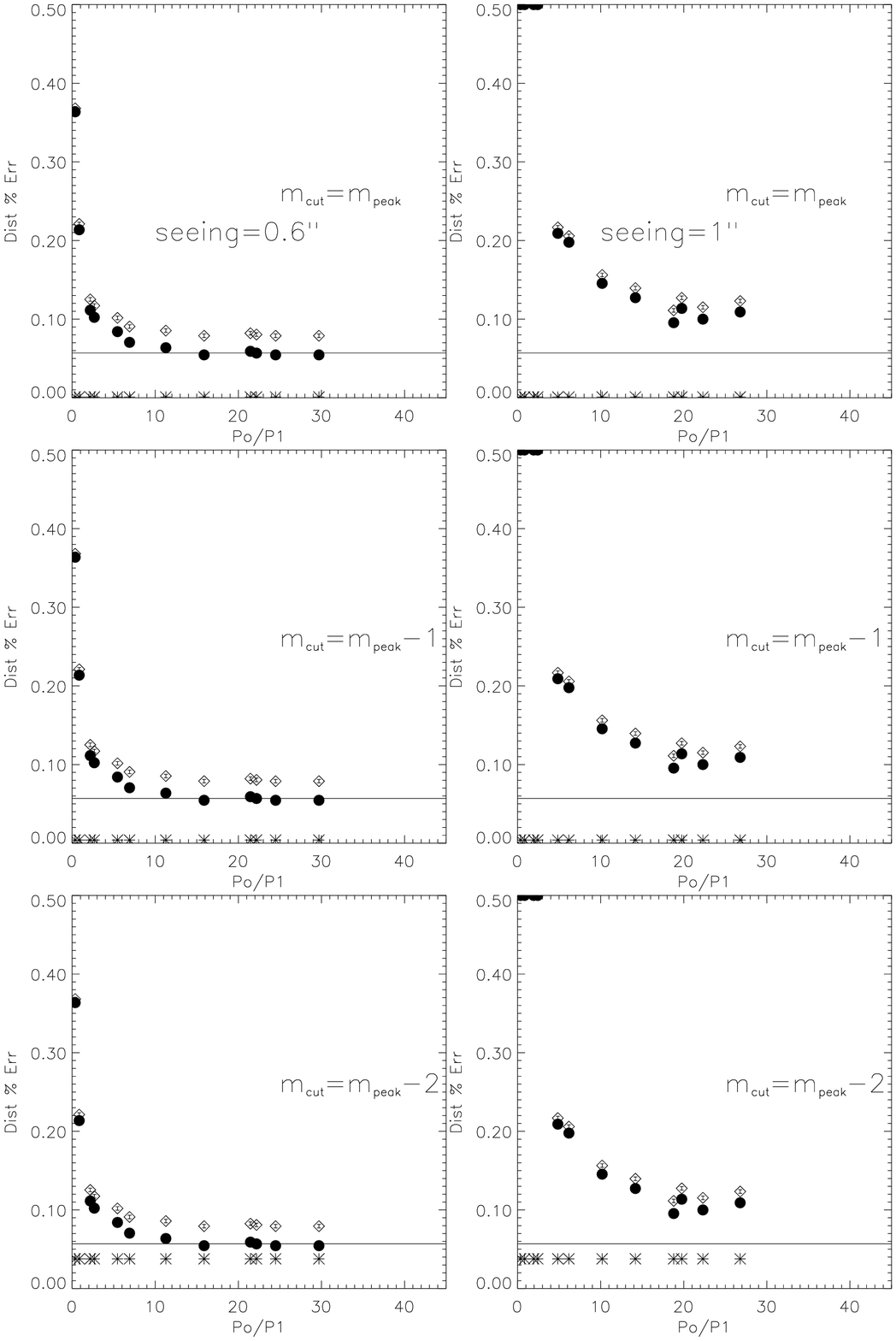,width=18cm,height=15cm}}

\caption{K--band error budget found from our simulations
for seeing = 0.6 \arcsec and 1 \arcsec.  The percentage error on distance 
is shown as a function of the signal--to--noise ratio. 
Diamonds represent the total error; filled circles, the error due only
to power spectrum fitting; and asterisks, the error due to external 
source detection and subtraction. The constant solid line 
is the percentage error induced by the zero point magnitude 
and SBF absolute magnitude.} \label{fig-perrK}
\end{figure*}

\noindent  Our results are plotted in Figure~\ref{fig-perrI} 
and Figure~\ref{fig-perrK}. 

The total variance of the power spectrum is the sum of the variance due to
the fluctuations and due to white noise. For a S/N $<$ 10, the white noise
dominates.  At higher S/N, the fluctuations dominate and 
we observe that the percentage error on the 
power spectrum reaches a plateau.  The plateau is
effectively reached after a signal--to--noise ratio of $\approx$ 10.

\noindent The error due to the presence of globular clusters 
and background galaxies was then considered.

\noindent In this case:

\begin{equation}
P_0=\sum \sigma^2_{SBF} +\sum \sigma^2_{gc} + \sum \sigma^2_{bg} \label{eq:fou2}
\end{equation}
\noindent and
\begin{equation}
P_1=\sum \sigma^2_{ph} +\sum \sigma^2_{RON}. \label{eq:fou3}
\end{equation}
\noindent The sum is taken over all non--zero pixels in the image, 
where $ \sigma^2_{gc}$ are fluctuations due to 
un--removed globular clusters, and $\sigma^2_{bg}$ represents
fluctuations due to un--removed background galaxies.

\noindent When performing SBF data reduction, the contribution 
from external sources was evaluated and subtracted by a two step process:

\noindent First, the external sources are detected and their 
magnitudes calculated. This allows us to build a luminosity 
function, complete down to a cut--off magnitude $m_{cut}$. 
The complete source luminosity function is then the sum of the 
globular cluster luminosity function (GCLF) plus the background galaxy 
luminosity function.

\noindent The globular cluster Gaussian luminosity function was 
assumed to be Gaussian: 
\begin{equation}
N(m) = \frac{N_{ogc}}{\sqrt{2\pi}\sigma} \ e^{\frac{-(m-m_{peak})^2}{2\sigma^2}}, \label{eq:gc3}
\end{equation}
\noindent  with $m_{peakV}=-7.11$ and $\sigma = 1.4 $\cite{har91}, 
and (V--I)=1 and (V -- K) = 2.23 \cite{jen98} for globular clusters.

\noindent $N_{ogc}$  is defined as:
\begin{equation}
N_{ogc} = S_{N} <gal> 10^{-0.4[M_{V}+15]},  \label{eq:sn}
\end{equation}
\noindent where $N_{ogc}$ is the mean number of globular clusters per pixel 
in the region, $<gal>$ the galaxy average flux in the region, $S_{N}$ the specific frequency,  
$M_{V}$ is the absolute magnitude of the region in the V--band. 
We considered $S_{N} = 2, 4 $ and $6$ \cite{har91,bla95}. 

\noindent For the background galaxies, we assume a 
power--law luminosity function \cite{ts88}
\begin{equation}
N(m) = N_{obg} 10^{\gamma(m-m_{og})} . \label{eq:bg}
\end{equation}
\noindent where  $N_{obg}$ is the mean number of galaxies per pixel. 
In the I-band we have assumed $\gamma = 0.27, m_{og} = 24.4,  
N_{obg} = 10^5 $ galaxies deg$^{-2}$ mag$^{-1}$ \cite{sma95}. 
In the K-band we have assumed  $\gamma = 0.30, m_{og} = 19,  
N_{obg} = 10^4 $ galaxies deg$^{-2}$ mag$^{-1}$ \cite{cow94}. 

\noindent The detected objects are masked from the image and the 
measurement of the fluctuation amplitude is done as described above.
The measured fluctuation amplitude will then contain the 
SBF amplitude plus the undetected external source fluctuations.

\noindent The second step in our SBF analysis then consists of 
estimating the residual external source fluctuations,
and their subtraction from the total power spectrum amplitude.
For the globular clusters, these residuals are given by \cite{bla95}:
\begin{eqnarray}
\sigma^2_{gc}=& \frac{1}{2} N_{ogc} 10^{0.8[m_0-m_{peak}+0.4\sigma^2ln(10)]}  \nonumber\\
& erfc[\frac{m_{cut}-m_{peak}+0.8\sigma^2ln(10)}{\sqrt{2}\sigma}] \label{eq:cgc}
\end{eqnarray}
\noindent Where $m_{0}$ is the telescope zero--point magnitude 
(the magnitude for which 1ADU/sec is collected) and $m_{cut}$ 
is the magnitude cut determined by the completeness of the sample 
in each region in which we measure SBF.
For the background galaxies, the residual contribution is given by :
\begin{equation}
\sigma^2_{bg}=\frac{N_{obg}p^2}{(0.8-\gamma) ln(10)}10^{0.8(m_0-m_{cut})-\gamma(m_{og}-m_{cut})}. \label{eq:cbg}
\end{equation}
\noindent where $p$ is the pixel scale.

\noindent The SBF amplitude is then estimated by:
\begin{equation}
\sigma^2_{SBF}= P_{o} - \Sigma \sigma^2_{gc}- \Sigma \sigma^2_{bg}. \label{eq:totsig}
\end{equation}
\noindent The error on the final estimate is the sum of the 
intrinsic error and the error from the external source power 
spectrum estimate.
In order to calculate the external source power spectrum, 
we must estimate the parameters $N_{ogc}$ and  $N_{obg}$. 
To this end, we fit the external source luminosity 
function down to $m_{cut}$ as:
\begin{equation}
N_{tot}=N_{ogc} e^{\frac{-(m-m_{peak})^2}{2\sigma^2}}+N_{obg} 10^{\gamma(m-m_{og})} \label{eq:ntottot}
\end{equation}
\noindent keeping $m_{peak}, \sigma^2$ and $m_{og}$ fixed 
during the fit.

\noindent In calculating the errors on $\sigma^2_{bg}$ and $\sigma^2_{gc}$ 
by error propagation theory, we assumed that the parameters with 
errors are $N_{ogc}$, $N_{obg}$, $m_0$, and $m_{peak}$ all the others 
being considered fixed. The errors are then given by:

\begin{equation}
\Phi= erfc[\frac{m_{cut}-m_{peak}+0.8 \sigma^2 ln(10)}{\sqrt{2}\sigma}]
\end{equation}

\begin{eqnarray}
\Delta^2(\sigma^2_{gc})=&  (\frac{1}{2} 10^{0.8[m_0-m_{peak}+0.4\sigma^2 ln(10)]})^2 *  \nonumber \\
& \{\Phi^2 \Delta^2(N_{ogc})+ \Phi^2 N_{ogc}^2 (0.8 ln (10))^2 \Delta^2(m_0) + \nonumber \\
&  N_{ogc}^2 (-0.8 ln (10) \Phi +  \sqrt{\frac{2}{\pi}}\frac{e^{-\Phi^2}}{\sigma})^2\Delta^2(m_{peak})  \} \nonumber \\
\end{eqnarray}

\begin{eqnarray}
\Delta^2(\sigma^2_{bg})=& \{\Delta^2(N_{obg})p^4 +(0.8 ln (10) N_{obg} p^2)^2 \Delta^2( m_0 )  \} \nonumber \\
&(\frac{10^{0.8(m_0-m_{cut})-\gamma(m_{og}-m_{cut})}}{(0.8-\gamma) ln(10)})^2.
\end{eqnarray}

%Figure--------------------

%--------------------------

%Figure--------------------

%--------------------------

%Figure--------------------
\begin{figure*}

\centerline{\psfig{file=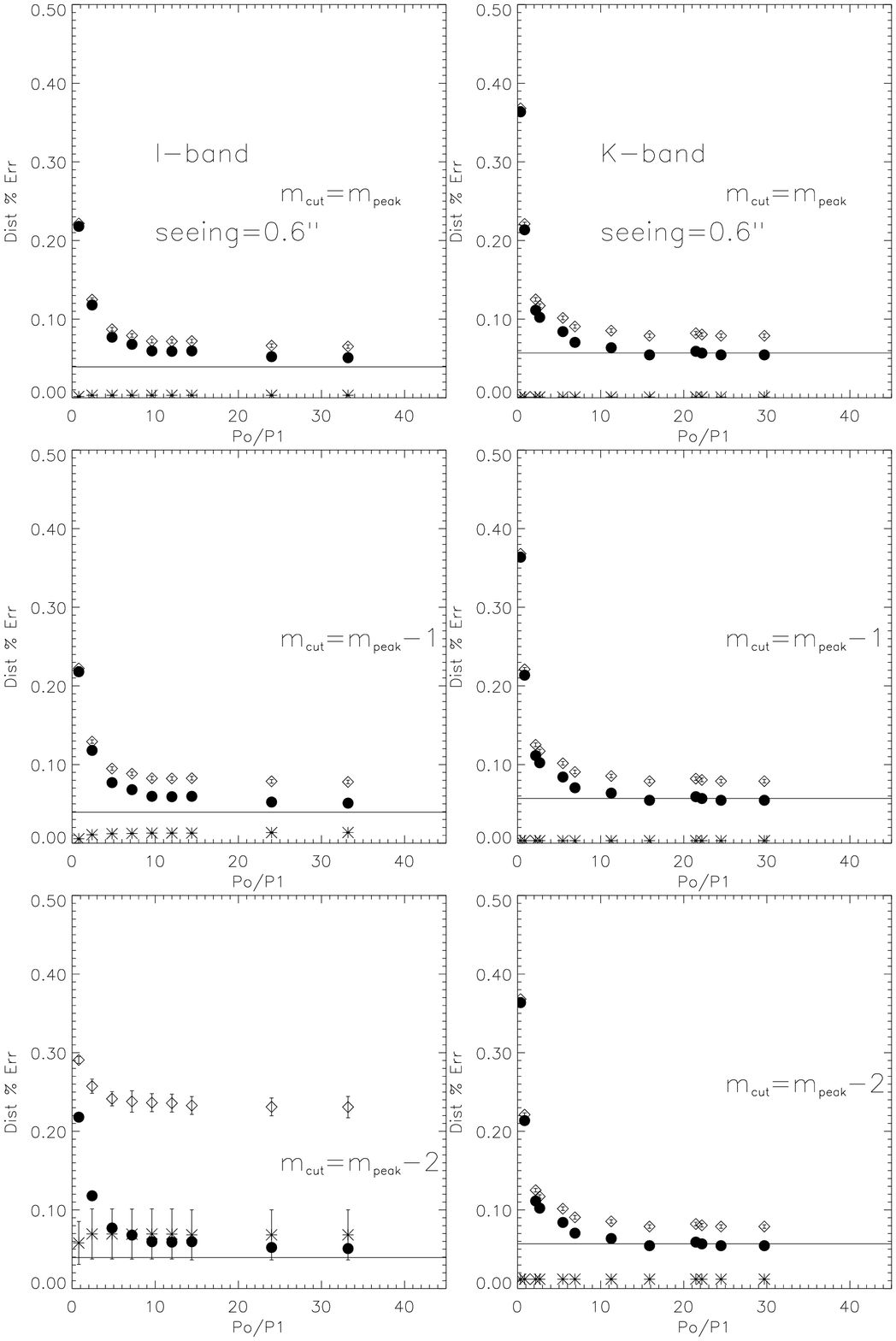,width=18cm,height=15cm}}

\caption{We consider here the case in which no error on the GCLF peak is included in the calculation. We plot I (on the left) and K--band (on the right) error budget found from our simulations
for seeing = 0.6$/arcsec$.  The percentage error on distance 
is shown as a function of the signal--to--noise ratio. 
Diamonds represent the total error; filled circles, the error due only
to power spectrum fitting; and asterisks, the error due to external 
source detection and subtraction. The constant solid line 
is the percentage error induced by the zero point magnitude 
and SBF absolute magnitude.} \label{fig-compIK}
\end{figure*}

%--------------------------

\noindent In I and K--bands, we simulated composite 
(globular cluster plus background galaxy) luminosity functions, 
assuming that the error for each $N$ as a function of magnitude 
is given by a Poisson distribution. We then fit these 
luminosity functions down to a completeness magnitude 
$m_{cut}$ for four hundred independent realizations, 
with the only free parameters being $N_{ogc}$ and  $N_{obg}$.
We calculated the errors on $N_{ogc}$ and  $N_{obg}$ from the
variance of their fitted values over the simulations.
Using these latter values, we then calculated the errors 
on the external source fluctuation residuals.
\noindent The error on the globular cluster peak has been taken to be 
0.15~mag when $m_{cut}$ is equal to $m_{peak}$, 0.20~mag when $m_{cut}$ is equal to $m_{peak}-1$, 0.25~mag when $m_{cut}$ is equal to $m_{peak}-2$. These values are an estimate of the error on the peak of the globular cluster luminosity function from published data \cite{fer00}. More detailed simulations are needed to better quantify this error. 

\noindent The errors due to zero--point determination and the absolute
magnitude calibration of the fluctuations were added to derive 
a total error on distance measurements.  We have taken the error in 
the zero--point magnitude to be 0.02~mag in the I--band, and 0.03~mag 
in the K--band. 
The error on the fluctuation absolute magnitude was set to 0.08~mag 
in the I--band, from the Tonry et al.(2000) calibration, and 0.12~mag 
from the dispersion of the K--band data available from Jensen et al. (1998).
Ferrarese et al. (2000) find a similar K--band dispersion of 0.11~mag.
Both these errors include the {\it random} error on Cepheid zero point, but
not  the {\it systematic} Cepheid zero point error of 0.16~mag \cite{fer00,mou00}, that was included only in our estimation of the total percentage error in Figure~\ref{fig-totperI} and  Figure~\ref{fig-totperK}.
If the errors on the zero-point magnitude and on the absolute magnitude calibration of the fluctuations would be the same in the K as in the I--band, the total percentage error on distance in the K--band will change by 0.01-0.03\%. 

\subsection{Results of the simulations}

\noindent Our results are presented as percentage 
errors on distance versus signal--to--noise ratio in 
Figure~\ref{fig-perrI} for the I--band and in Figure~\ref{fig-perrK} 
for the K--band; seeing equal to 0.6$\arcsec$ and 1$\arcsec$ were
considered in both cases. 
The error budget shown in Figure~\ref{fig-perrI} and Figure~\ref{fig-perrK}
does not include the systematic uncertainty on the Cepheid zero point. Diamonds show the total error due to power spectrum fitting, 
external source detection and subtraction, plus errors 
in zero--point magnitude and SBF absolute magnitude calibration.
Filled points show the relative contribution to the error budget 
from power spectrum fitting; asterisks, the contribution due to 
external source detection and subtraction; and the constant, solid line,
the contribution of zero--point magnitude and SBF absolute magnitude 
calibration uncertainties.
From top to bottom in a column, the completeness magnitude $m_{cut}$
for external source detection decreases and is expressed relative
to the peak of the globular cluster luminosity function.

\noindent The percentage distance error is averaged over five 
different annuli with a width of $\approx 12$~$\arcsec$. 
The error bar is given by the standard deviation over these five annuli.
The assumed $S_{N}$ in the plot is 6.
For lower frequencies, $S_{N}$=2 and 4, the total error does 
not change dramatically.
%Figure--------------------

\begin{figure*}

\centerline{\psfig{file=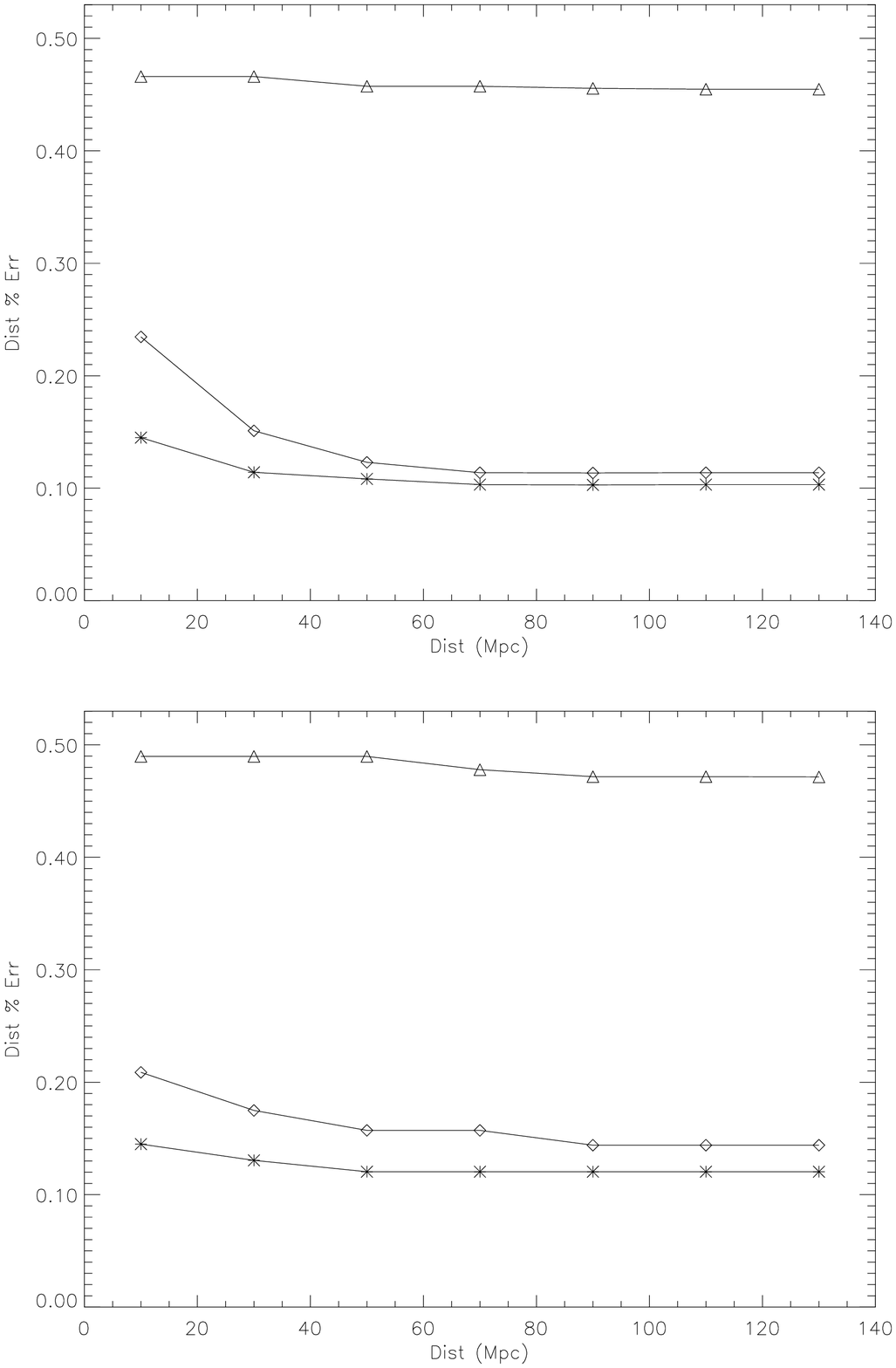,width=12cm,height=12cm}}

\caption{Total I--band distance percentage error versus distance 
calculated for signal--to--noise ratios corresponding to integration 
times that permit a detection of external sources at 90\% completeness 
down to $m_{cut}$ equal to the peak magnitude $m_{peak}$ of the globular 
cluster distribution (asterisks), $m_{peak}$-1 (diamonds), 
and $m_{peak}$-2 (triangles). The top panel is for a seeing 
equal to 0.6\arcsec, and bottom one for a seeing equal to 
1 \arcsec} \label{fig-totperI}

\end{figure*}

%--------------------------

%Figure--------------------

\begin{figure*}

\centerline{\psfig{file=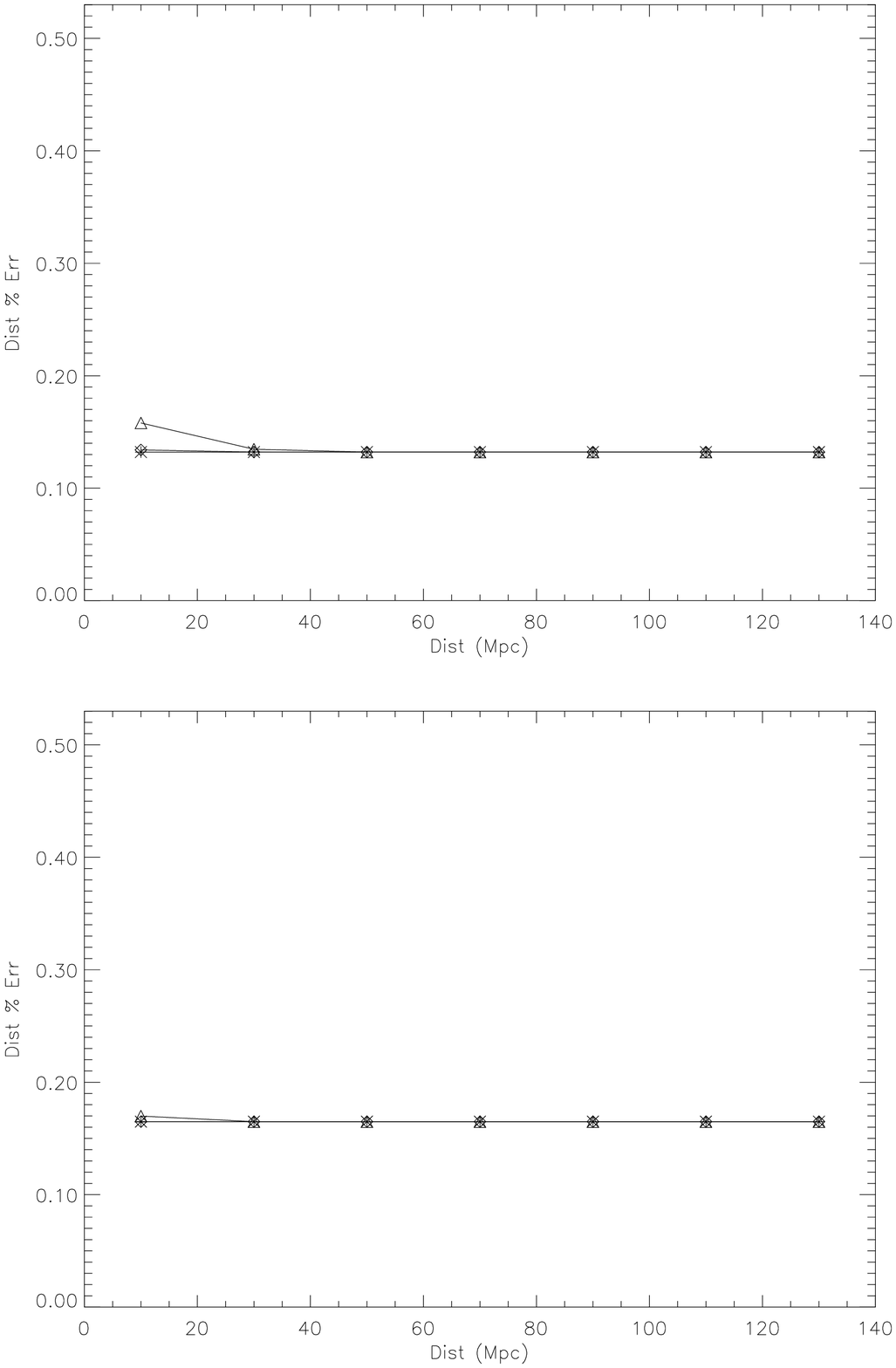,width=12cm,height=12cm}}

\caption{Total K--band distance percentage error versus distance 
calculated for signal--to--noise ratios corresponding to integration 
times that permit a detection of external sources at 90\% completeness 
down to $m_{cut}$ equal to the peak magnitude $m_{peak}$ of the globular 
cluster distribution (asterisks), $m_{peak}$-1 (diamonds), 
and $m_{peak}$-2 (triangles). The top panel is for a seeing 
equal to 0.6\arcsec, and bottom one for a seeing equal to 
1 \arcsec} \label{fig-totperK}

\end{figure*}

%--------------------------

%Figure--------------------

\begin{figure*}

\centerline{\psfig{file=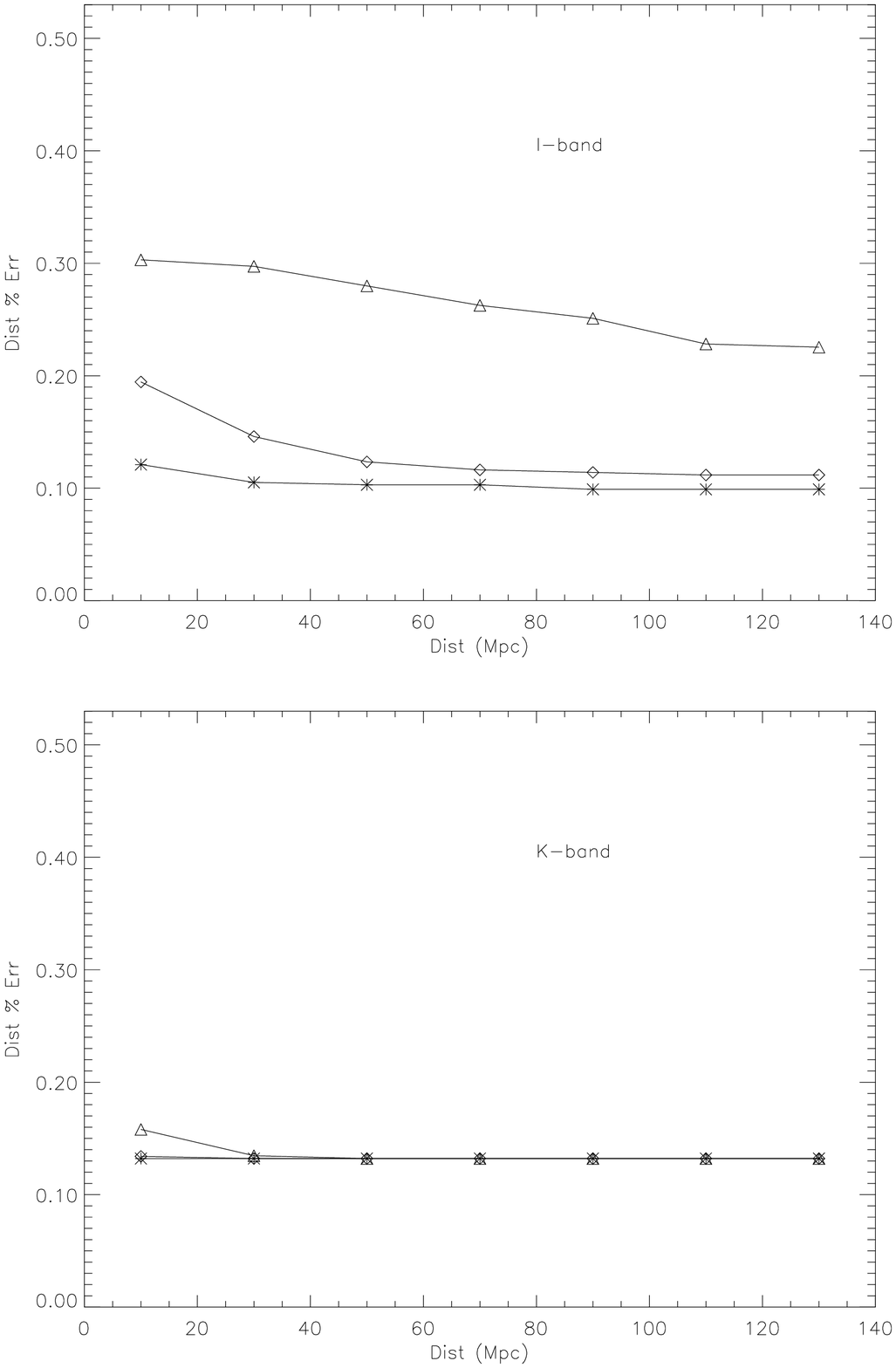,width=12cm,height=12cm}}

\caption{We consider here the case in which no error on the GCLF peak is included in the calculation. From top to bottom, we plot I and K--band total distance percentage error versus distance 
calculated for signal--to--noise ratios corresponding to integration 
times that, in each band,  permit a detection of external sources at 90\% completeness 
down to $m_{cut}$ equal to the peak magnitude $m_{peak}$ of the globular 
cluster distribution (asterisks), $m_{peak}$-1 (diamonds), 
and $m_{peak}$-2 (triangles). The  seeing is 
equal to 0.6\arcsec.} \label{fig-totcompIK}

\end{figure*}

%--------------------------

\noindent We plot in Figures~\ref{fig-totperI} and \ref{fig-totperK} 
the total percentage distance error as a function of distance 
calculated for signal-to-noise ratios that will permit us to obtain a completeness of 90\% on 
external source detection for different cut-off magnitudes $m_{cut}$.
In this percentage error the systematic error on Cepheid zero point, 0.16~mag  \cite{fer00,mou00},
was included.
As we can see from the figures, with a seeing equal to 0.6$\arcsec$, 
in both I and in K--band it is possible to measure distance 
with an accuracy of 10\% if external sources are detected down 
to one magnitude brighter than the peak magnitude of the globular 
cluster luminosity function. 
With a seeing equal to 1$\arcsec$, for large distances, the typical errors on
 measurements are similar.
The key quantity is therefore the integration time necessary for the detection of
external sources (which depends both on distance and on seeing) and crowding
effects.

\noindent Reaching  the peak magnitude of the globular 
cluster luminosity function is time expensive to realise in practice.
Typical observations will attain  a completeness in external source detection
between one to one and half magnitude fainter than the estimated peak magnitude of the
globular cluster luminosity function. This will be the major source of error
on large distance measurements.
In our simulations, we fixed the peak and the dispersion of the globular cluster luminosity function, when fitting it. We assumed an error on the globular cluster luminosity function peak equal to 0.15~mag when $m_{cut}$ is equal to $m_{peak}$, 0.20~mag when $m_{cut}$ is equal to $m_{peak}-1$, 0.25~mag when $m_{cut}$ is equal to $m_{peak}-2$. These values have been chosen from published data \cite{fer00} and they take into account the fact that the error will increase as $m_{cut}$ falls below $m_{peak}$. More detailed simulations would be needed to further quantify this point.
We show in Figure~\ref{fig-compIK} and Figure~\ref{fig-totcompIK} what 
the calculated errors would be in both bands when not taking in account any error on $m_{peak}$, in the case in which the seeing is equal to 0.6$\arcsec$.

\noindent We can now compare the two bands. From Figure~\ref{fig-totperI} 
and  Figure~\ref{fig-totperK}, the first difference to note between
the two bands is that, while in the I--band the percentage error 
changes significantly as a  function of the cut--off magnitude, 
in the K--band the error tends to degenerate to closest values. 
This is because of the higher color contrast of the fluctuations in 
K with respect to the external source color in this band. 
The percentage contribution of external source fluctuations 
to SBF fluctuations does not change as dramatically as in the I--band 
when the cut--off changes by one or two magnitudes. One could at this 
point think that this peculiar characteristic of the K--band 
would permit us to reach larger distances in this band than in 
the I--band. This is actually not true from the ground, due to the 
long integration time required to detect external sources 
down to a significant cut--off magnitude in the presence
of the high background in the K--band.
%A seeing equal to 0.6 $\arcsec$ is expected at Paranal with a probability of 30 \% in the I--band and of 50 \% in the K--band.

\noindent In fact, given these results and knowing the 
VLT I and K--band integration times required to obtain the signal--to--noise 
ratios that we considered, we can put limits on the distance 
attainable within a given percentage distance error.
For a galaxy of absolute V magnitude of -23,
we have predicted from simulations that it is possible in the I--band, with a total (integration time plus overheads) exposure time no 
larger than 12 hours and a sky of 19.2 mag/arcsec$^2$, to reach distances up to $\approx 90$ Mpc, 
or, for $H_{0}$=70 km/s/Mpc, $\approx$ 6500 km/s, 
within an error of $\approx$ 10\% for a seeing equal to 0.6 $\arcsec$ (assuming to detect external source up to a cutoff magnitude one magnitude below the peak of the globular cluster luminosity function). 
 If the total exposure time can be increased to 24 hours, it will be possible, under the same conditions, to reach distances around $\approx$ 10000 km/s.        In the K--band, with a total exposure time no larger than 12 hours, 
it is possible to reach distances up to $\approx 40$ Mpc, or, for 
$H_{0}$=70 km/s/Mpc, $\approx$ 3000km/s, with an error of around
10 \% and a seeing equal to 0.6$\arcsec$ when a sky of 13 mag/arcsec$^2$ is assumed; or  $\approx$ 3500km/s when a  sky magnitude of 13.7 mag/arcsec$^2$ is assumed (assuming to detect external source up to a cutoff magnitude two magnitudes below the peak of the globular cluster luminosity function). If the total exposure
time can be increased to 24 hours, it will be possible, under the 
same conditions, to reach distances around $\approx$ 4000 km/s 
with an error around 15 \%, with a seeing equal to 1$\arcsec$.
We have assumed an overhead of approximately 50\%.

\noindent We thus conclude that while I--band SBF observations 
at ground--based large telescopes will permit one to extend 
the current ground--based sample out to 6500--10000~km/s, 
K--band SBF measurements from the ground alone will be 
limited by the high infrared background.

\noindent Hybrid approaches, as already pointed out by Jensen and 
his collaborators \cite{jen96,jen98,jen99,jen00} can combine SBF observations 
in the K--band with external source detection in the I--band, 
reducing expensive exposure time in the K--band while maintaining a correct
estimation of the external source contribution. 
On the one hand, when external sources are detected in the I--band around 
one magnitude below $m_{peak}$, this approach has the advantage 
of providing two independent (even if there will not be an independent
external source detection) SBF distance measurements, or of
providing information on the galaxy stellar population.
On the other hand, as can be seen from Figure~\ref{fig-totperK}, when external sources are detected for $m_{cut}$ larger than $m_{peak}$ - 1, 
I--band distance measurements present errors from 10$\%$ to 50$\%$, 
while K--band measurement errors at the same $m_{cut}$ are around $10\%$.
A more detailed study of combining I and K--band surveys has to be explored, in order to define the optimal approach depending on distance and on observing 
conditions.

\noindent We must point out that the errors that we have derived  
are lower limits. We are estimating errors within the available  
observational constraints and standard data processing.
We have not considered other sources of error, such as anomalous 
boxiness or diskiness of the galaxy, or profiles differing  
from a standard de Vaucouleurs profile.
We have assumed that these effects are negligible 
and corrected by the smooth fit of the residual image 
after the first subtraction of the galaxy model.
We did not consider crowding effects on the detection of external sources.
 We assumed the error on the globular cluster luminosity function peak in function of $m_{cut}$, but did not strengthened these values with simulations. More detailed simulations would be needed to further quantify these points.

\subsection{Comparison with VLT I--band observations}

\noindent We can compare these theoretical predictions with VLT I-band SBF
that we measured in IC 4296 \cite{mei00}, an elliptical galaxy in  Abell 3565
($\approx$ 3500 km/s). For this galaxy we have measured  a distance modulus of
$(\overline I_{o,k} - \overline M_I) = 33.44 \pm 0.17$ mag and a galaxy
distance of $49 \pm 4$ Mpc. From HST I--band SBF measurements, Lauer et
al. (1998) have obtained  $(\overline I_{F814} - \overline M_{F814}) = 33.47
\pm 0.13$ mag with a derived distance of 49 $\pm$ 3 Mpc. The seeing of this
observation was 0.7$\arcsec$. We obtain a completeness magnitude at around one
magnitude brighter than the globular cluster luminosity function peak. The percentage error
on distance that we obtained on IC~4296 is $\approx$ 8\%, This error is
consistent with our Monte Carlo prediction of $\approx$10\% for external
source detection up to one magnitude brighter than the globular cluster
luminosity function peak, as  presented in section 3.2, Fig. 6.
 The error on distance that we obtain with
2.3 hours of integration time are comparable to 3.2 hours integration time from the Hubble Space Telescope \cite{lau98}. This result confirms the potential of 8--m class telescope in large distance ground--based SBF observations.
 Ground--based observations from large telescopes are thus competitive with space observations, but  with the advantage of larger fields of view, permitting 
one to measure  more ellipticals in the same field. Our conclusions are also in agreement with the recent observations and conclusions from Tonry et al. (2001).

\section{Discussion}

\noindent We have evaluated in this paper the potential of SBF measurements in the I and in the K--band. The method is limited by stellar population effects and by the errors that come from observational conditions and data treatment.
The I--band is currently well-calibrated by the Tonry et al.(2000) linear relation between SBF amplitudes and galaxy (V--I) color. We have studied the effect of stellar population on the K--band amplitudes with Bruzual and Charlot stellar population model predictions.

\noindent Bruzual and Charlot (2000) single
age -- single metallicity models with a Salpeter IMF and  solar metallicity,
predict an average $\overline M_K=-5.48 \pm 0.22$ mag and (V--I) colors
between 1.1 and 1.39. This value is consistent
with previous  K--band observations in the Virgo and  Fornax cluster by Jensen
et al. (1998). The average and standard deviation of the Jensen et al. (1998)
sample is $\overline M_K=-5.61 \pm 0.12$ mag for (V--I) colors between 1.15
and 1.27.

\noindent We have also examined the prediction of Bruzual and Charlot  models with a Salpeter IMF in the case of composite populations.

\noindent These models (cf. Liu et al. 2000) predict that the presence of young populations
and high metallicities will increase the amplitude of the fluctuations, while low
metallicities  will lower their value.

\noindent Independent models by Blakeslee et al. (2001)also predict higher amplitude fluctuations in younger populations,
but a much weaker metallicity dependence.

\noindent The variation of the absolute SBF might explain the  higher than
average SBF, observed in some of the low signal--to--noise sample of Jensen et
al. (1998) and Pahre \& Mould (1994), as well as the higher than standard K--band
fluctuations observed by Mei et al. (2001a) in NGC~4489. At present, galaxies
with anomalous low fluctuations have not been observed.
 NGC~4489 has lower (V--I) than the Jensen et al. (1998) sample from which
the current K~--~band calibration is derived. From theoretical predictions, its
higher K--band SBF amplitude can be explained both by the presence of an
intermediate age population (from Liu et al. 2000 predictions) or by a
difference in metallicity, quantified from the difference in (V--I)
(predictions from Blakeslee et al. 2001 models). 

\noindent Additional K--band SBF data over a larger range of galaxy colors and
luminosities are necessary to improve the
precision on the empirical calibration. A better calibration at lower (V--I)
would permit one to
discriminate between the Liu et al. (2000) and Blakeslee et al. (2001)
predictions.  

\noindent We have also quantified the limit of the method in both the I and
the K--band by studying the error budget that comes from observational conditions
and data treatment, using Monte Carlo simulations. In our simulation K--band
galaxies were considered over the range of the Jensen et al. (1998) calibration,
because measurements on large distances will be based on luminous, red galaxies.

\noindent According to our results, the VLT and other very large telescopes will permit us to extend SBF distance measurements in the I--band to a distance $\approx$ 6500 km/s within a percentage error on single galaxy distance of $\approx$ 10 \%.  In the K--band, distances $\approx$ 3000 km\/s will be reached with a percentage error on distance of $\approx$ 10 \%.  
This error can be compared to the single galaxy distance percentage error $\approx$ 10 \% that can be obtained by the PNLF method and Supernovae Type Ia, and it is smaller than the  15-20\% distance error that can be obtained by Tully-Fisher and $D_{n}-\sigma$.

\noindent This means that SBF distance measurements on large telescopes will
permit us to extend the current ground--based I--band  sample to twice the
current distance limit, to calculate bulk flows beyond the current limit of 3000 km/s \cite{ton00}, and to compare distances and peculiar velocities with those derived from other methods.

\noindent On the other hand, measuring in the K--band alone does not permit us
to go very deep in distance from the ground, due the high background level in
the infrared and the high integration times needed to detect external sources.
Jensen et al. (1998) have made extensive use of optical images for external source detection. They have derived globular cluster luminosity functions from the I--band image of the galaxies that they observed in the K--band, and employed it to correct the external source contribution to the SBF fluctuations in the K--image. This method requires both I and K images of the same galaxies, and permits one to clearly identify the SBF fluctuations in the K--band with no contamination.

\noindent The optimal use of both bands for distance
measurements has to be studied with
more detailed simulations, but it is nevertheless useful in two other respects. Firstly, it enables us to discriminate among stellar population models. Secondly, the
use of combined I and K--band SBF will help us  gain knowledge on K--band
SBF calibration in preparation for space--based observations, that are possible on NICMOS on the HST \cite{jen00} and will be possible with the NGST. In fact, in space the background in the K--band is significantly lower and the required source extraction will be possible in reasonable integration times.

\noindent Current K--band measurements from the ground might thus help in
stellar population model discrimination and study of stellar population age and
metallicity in elliptical galaxies \cite{liu00,bla01}, and can lay the basis for further K--band SBF measurements from the space.

\section{Summary}
\noindent In this paper we have shown that: 

\begin{enumerate}

\item Recent stellar population models \cite{bla01,liu00} give contradictory
  predictions for K--band SBF. It is necessary to enlarge the sample of K--band measurements
  to lower (V--I) to obtain a dependable calibration of the K--band SBF
  absolute magnitude. At the same time, the Jensen et al. (1998) empirical
  mean is compatible with theoretical models for bright, red galaxies
  that will be the galaxies observed at large distances. In our simulations we
  used Bruzual \& Charlot (2000) stellar population models, as did Liu et
  al. (2000). 
 There models predict that K-band SBF are driven by both age and
 metallicity, in the sense that increasing metallicity and decreasing
 age produce larger K--band SBF (i.e. brighter SBF magnitudes).  The
 effect of metallicity is much larger than the effect of age.
 The K--band SBF amplitude becomes higher for redder populations.
       It seems
 possible that stellar population effects will have a larger impact in
 the K-band, especially for lower luminosity ellipticals, making
 individual K-band based distance measurements more uncertain.  In
 other words, there may not be a a simple linear relationship between
 color and K-band SBF magnitude, as there is in the I-band.  Further
 K-band SBF measurements of selected galaxy samples remain critically
 necessary for understanding these stellar population effects and for
 disentangling the current disagreements between theoretical
 predictions of K-band SBF magnitudes and colors.  For this kind of
 study, ground--based work is appropriate.  Ground--based observations in the K--band will lay the basis for more precise K--band  SBF measurements from space, for example, with NICMOS on HST \cite{jen00} and, in the future, with the Next Generation Space Telescope (NGST).

\item Our Monte Carlo simulations indicate that very large ground--based
  telescopes, such as the VLT,  will permit one to extend I--band SBF distance measurements out to 6000--10000 km/s, with an error of 10--15\%. 

\item  Relative to ground--based I-band SBF measurements, it appears that, unless optical data are available to aid in the external source removal,
  ground--based K-band SBF measurements are prohibitively expensive for
  obtaining large numbers of distance measurements at large distances, while
  space--based K--band observations are more efficient.

\end{enumerate}

These conclusions need to be strengthened by exploring a wider range of parameters and by more extensive simulations of errors due to the corrections of the external source contribution, to the PSF template fitting, and to irregular isophotes. Our approach is novel for its application of realistic Monte Carlo simulations of SBF observations. It will be further developed in subsequent work.
\begin{acknowledgements}

We are grateful to S. Charlot for proving Bruzual \& Charlot  models and to our referee, John Blakeslee, for his useful comments. S. Mei thanks M. Romaniello and P.A. Duc for the interesting discussions, J.G. Bartlett for reading the manuscript, and acknowledges support from the European Southern Observatory Studentship programme and Director General's Discretionary Fund.

\end{acknowledgements}

\end{document}